\documentclass{article}
\def\Ref#1{(\ref{#1})}
\usepackage{amsmath}
\begin{document}
\begin{titlepage}
\begin{center}
{\large \bf Dynamical phase transition of a one-dimensional
kinetic Ising model with boundaries} \vskip 2\baselineskip
\centerline {\bf Mohammad Khorrami
$^{a}$\footnote{e-mail:mamwad@iasbs.ac.ir}
 {\rm and}
 Amir Aghamohammadi $^{b}$ \footnote {e-mail:mohamadi@theory.ipm.ac.ir}}
 \vskip 2\baselineskip
{\it $^a$ Institute for Advanced Studies in Basic Sciences,}
\\ {\it P. O. Box 159, Zanjan 45195, IRAN}\\
{\it $^b$ Department of Physics, Alzahra University, Tehran 19384,
IRAN}
\end{center}


\begin{abstract}
\noindent The Glauber model on a one-dimensional lattice with
boundaries (for the ferromagnetic- and anti-ferromagnetic case) is
considered. The large-time behaviour of the one-point function is
studied. It is shown that, for any positive temperature, the
system shows a dynamical phase transition. The dynamical phase
transition is controlled by the rate of spin flip at the
boundaries, and is  a discontiuous change of the derivative of the
relaxation time towards the stationary configuration.
\end{abstract}
\end{titlepage}
\newpage
\section{Introduction}
The principles of equilibrium statistical mechanics are well
established. But, thermal equilibrium is a special case, and
little is known about the properties of systems not in
equilibrium, for example about the  relaxation toward the
stationary state. Some interesting problems in non-equilibrium
systems are non-equilibrium phase transitions described by
phenomenological rate equations, and the way the system relaxes to
its steady state. As mean-field techniques, generally, do not give
correct results for low-dimensional systems, people are motivated
to study exactly-solvable stochastic models in low dimensions.
Moreover, solving one-dimensional systems should in principle be
easier. Exact results for some models on a one-dimensional lattice
have been obtained, for example in [1--14]. Different methods have
been used to study these models, including analytical and
asymptotic methods, mean field methods, and large-scale numerical
methods.

The Glauber dynamics was originally proposed to study the
relaxation of the Ising model near equilibrium states. It was also
shown that, there is a relation between the kinetic Ising model at
zero temperature and the diffusion annihilation model in one
dimension. There is an equivalence between domain walls in the
Ising model and particles in the diffusion annihilation model.
Kinetic generalizations of the Ising model, for example the
Glauber model or the Kawasaki model, are phenomenological models
and have been studied extensively [15--20 ]. Combination of the
Glauber and the Kawasaki dynamics has been also considered
[21--23].

In \cite{MA}, an asymmetric generalization of the zero-temperature
Glauber model on a lattice with boundaries was introduced. It was
shown there that, in the thermodynamic limit, when the lattice
becomes infinite, the system shows two kinds of phase transitions.
One of these is a static phase transition, the other a dynamic
one. The static phase transition is controlled by the reaction
rates, and is a discontinuous change of the behavior of the
derivative of the stationary magnetization  at the end points,
with respect to the reaction rates. The dynamic phase transition
is controlled by the spin flip rates of the particles at the end
points, and is a discontinuous change of the relaxation time
towards the stationary configuration. Other generalizations of the
Glauber model consist of, for example, alternating-isotopic chains
and alternating-bound chains (see \cite{GO}, for example). People
have also considered phase transitions induced by boundary
conditions (see [26--28], for example).

The scheme of the paper is as follows. In section 2, the model is
introduced, the rates are determined using the detailed balance,
and the steady state configuration of the magnetization is
obtained. In section 3, the dynamical phase transition of the
system is investigated, and it is shown that it does show a
dynamical phase transition, provided the temperature of the system
is not zero.

\section{Kinetic Ising model on a one-dimensional lattice with boundaries}
The model being addressed, is the Glauber model on a
one-dimensional lattice with boundaries. In the Glauber model, the
interaction is between three neighboring  sites. Spin flip brings
the system into equilibrium with a heat bath at temperature $T$. A
spin is flipped with a rate $\mu = 1 - \tanh ({ 2\beta J})$
whenever the spin of both of its neighboring sites are the same as
it and is flipped with a rate $\lambda = 1 + \tanh ({ 2\beta J})$
whenever the spin of both of its neighboring sites are in the
opposite direction. (Here $\beta :=1/(KT)$.) At domain boundaries,
spins are flipped with unit rate. So the interactions can be
written as,
\begin{align}
\uparrow \; \uparrow \; \uparrow \; \to \; \uparrow \; \downarrow \;
 \uparrow \; \mbox{ and }
\downarrow \; \downarrow \; \downarrow \; \to \; \downarrow \;  \uparrow
\; \downarrow \; & \mbox{with rate} & \mu \nonumber \\
\uparrow \; \downarrow \; \uparrow \; \to \; \uparrow \; \uparrow
\; \uparrow \; \mbox{ and } \downarrow \; \uparrow \; \downarrow \; \to
 \; \downarrow \;  \downarrow \;
\downarrow \; & \mbox{with rate} & \lambda \nonumber \\
\uparrow \; \uparrow \; \downarrow \;
\rightleftharpoons \;
\uparrow \; \downarrow \; \downarrow \;
\mbox{ and }
\downarrow \; \downarrow \; \uparrow \;
\rightleftharpoons \; \downarrow \;  \uparrow \;
\uparrow \; & \mbox{with rate} & 1 \nonumber
\end{align}

Consider a lattice with $L$ sites and  the Glauber dynamics as the
interaction. The spin of the first site may flip with the
following rates
\begin{align}
\uparrow \; \downarrow  \to \;  \downarrow \; \downarrow  & \mbox{with rate} & g_1 \nonumber \\
\uparrow \; \uparrow  \to \;  \downarrow \; \uparrow  & \mbox{with rate} & g_2 \nonumber \\
\downarrow \; \uparrow  \to \;  \uparrow \; \uparrow  & \mbox{with rate} & g_3 \nonumber \\
\downarrow \; \downarrow  \to \;  \uparrow \; \downarrow  &
\mbox{with rate} & g_4, \nonumber
\end{align}
and the spin of the last site may flip with the following rates
\begin{align}
\downarrow \; \uparrow  \to \;  \downarrow \; \downarrow  & \mbox{with rate} & h_1 \nonumber \\
\uparrow \; \uparrow  \to \;  \uparrow \; \downarrow  & \mbox{with rate} & h_2 \nonumber \\
\uparrow \; \downarrow  \to \;  \uparrow \; \uparrow  & \mbox{with rate} & h_3 \nonumber \\
\downarrow \; \downarrow  \to \;  \downarrow \; \uparrow  &
\mbox{with rate} & h_4. \nonumber
\end{align}
It is known that the time evolution equation for the one-point
functions in the bulk are expressed in terms of only the one-point
functions \cite{RG}. To make this true for the boundaries as well,
the following relations should hold.
\begin{align}\label{cl}
g_1+g_4=& g_2+g_3\nonumber\\
h_1+h_4=& h_2+h_3.
\end{align}

One may give a physical meaning to the parameters $g_i$ and $h_i$,
by demanding the detailed balance to hold. Consider the energy
${\mathcal E}$ of the system to be
\begin{equation}
{\mathcal E} =-(B_1 s_1+B_L s_L+J\sum_{i=1}^{L-1} s_i s_{i+1}),
\end{equation}
then, the detailed balance demands
\begin{align}\label{db}
&\mathcal{R}(s_1s_2\to s'_1s_2)\exp\{\beta (B_1s_1+Js_1s_2+\cdots
)\}\nonumber\\
= &\mathcal{R}(s'_1s_2\to s_1s_2)\exp\{\beta
(B_1s'_1+Js'_1s_2+\cdots )\},
\end{align}
where $\mathcal{R}(s_1s_2\to s'_1s_2)$ is the rate of  the spin
flip of the first site from $s_1$ to $s'_1$. Equation (\ref{db})
shows that
\begin{equation}\label{fs}
\mathcal{R}(s_1s_2\to s'_1s_2)=f(s_2)\exp\{-\beta s_1(B_1+Js_2)\}.
\end{equation}
The exponential term in the above equation is at most linear in
terms of $s_1$. So,
\begin{equation}\label{fbs}
\mathcal{R}(s_1s_2\to s'_1s_2)=\bar f(s_2)[1-s_1\tanh \beta
(B_1+Js_2)].
\end{equation}
Then,
\begin{align}\label{gggg}
g_1&=\bar f(-1)[1-\tanh \beta (B_1-J)]\nonumber \\
g_2&=\bar f(1)[1-\tanh \beta (B_1+J)] \nonumber\\
g_3&=\bar f(1)[1+\tanh \beta (B_1+J)] \nonumber\\
g_4&=\bar f(-1)[1+\tanh \beta (B_1-J)].
\end{align}
The condition of exact solvability (\ref{cl}) (the closure of time
evolution equation of one-point functions) leads to
\begin{equation}\label{fbar}
\bar f(1)=\bar f(-1).
\end{equation}
This means that the {\it inertia} of the first spin against
spin-flip does not depend on the second spin. A similar expression
can be written for the rate of the spin-flip of the last site.

For the infinite lattice, the Glauber model has a particle
reaction-diffusion interpretation. If the spins of the neighboring
sites are different (at a domain wall), one may consider  the link
between that sites as a particle. When the spins of the
neighboring sites are the same (no domain wall), one may consider
the link between the sites as a vacancy. Then the Glauber model
turns into a reaction-diffusion model:
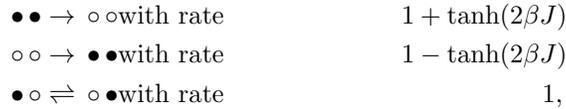
\begin{align}
\bullet\, \bullet \to \, \circ\, \circ &\mbox{with rate} & 1 + \tanh ({ 2\beta J}) \nonumber \\
\circ\, \circ \to \, \bullet\, \bullet & \mbox{with rate} & 1 - \tanh ({ 2\beta J}) \nonumber \\
\bullet\, \circ \rightleftharpoons \, \circ\, \bullet & \mbox{with
rate} & 1,\nonumber
\end{align}
where a particle (a vacancy) is denoted by $ \bullet$ ($\circ$).
For the Glauber model with boundaries, to have a consistent
particle model, one has to impose
\begin{align}
g_1=g_3\qquad  & g_2=g_4 \nonumber \\
h_1=h_3\qquad  & h_2=h_4.
\end{align}
Then, the injection and extraction of particle at the first site
are
\begin{align}
\bullet\, \to \circ  &\mbox{with rate} & g_1=g_3 \nonumber \\
\circ \to \, \bullet & \mbox{with rate} & g_2=g_4,
\end{align}
and the injection and extraction of particles at the last site are
\begin{align}
\bullet\to \, \circ &\mbox{with rate} & h_1=h_3 \nonumber \\
\circ \to \, \bullet &\mbox{with rate} & h_2=h_4,
\end{align}
Now, consider the general case where only the conditions
(\ref{cl}), which guarantee the closure of the time evolution, are
satisfied. We have
\begin{align}\label{2}
\langle\dot s_k\rangle =&-2\langle s_k\rangle +
 (\langle s_{k+1}\rangle +\langle s_{k-1}\rangle )
 \tanh ({ 2\beta J})\qquad 1<k<L \nonumber\\
\langle\dot s_1\rangle =& -(g_2+g_3)\langle s_1\rangle + (g_1-g_2)\langle s_2\rangle
+(g_3-g_1)\nonumber\\
\langle\dot s_L\rangle =& -(h_2+h_3)\langle s_L\rangle + (h_1-h_2)\langle s_{L-1}\rangle
+(h_3-h_1).
\end{align}
The steady-state solution to (\ref{2}) is
\begin{equation}\label{4}
\langle s_k\rangle =D_1z_1^k+D_2z_2^{k-L-1},
\end{equation}
where
\begin{equation}\label{5}
z_1=z_2^{-1}=\tanh (\beta J).
\end{equation}
It can be shown that in the thermodynamic limit ($L\to \infty$),
\begin{align}
D_1={g_1-g_3\over (g_1-g_2)z_1^2-(g_2+g_3)z_1}  \nonumber \\
D_2={h_1-h_3\over (h_1-h_2)z_1^2-(h_2+h_3)z_1}. \nonumber
\end{align}
$D_1$, and $D_2$ are continuous functions of the rates. So the
behavior of $\langle s_k\rangle$ near the ends of the lattice
varies continuously with rates, and there is no phase transition.

\section{The dynamical phase transition of the system}
The average magnetization per site $m(t)$ is
\begin{equation}\label{m}
m(t)={1\over L}\sum_{k=1}^L\langle s_k(t)\rangle .
\end{equation}
In the thermodynamic limit, the boundary terms are negligible, and
\begin{equation}\label{dm}
  {{\rm d}\over {\rm d}t}m(t)=2[\tanh(2\beta J)-1]m(t).
\end{equation}
Then, the same as Glauber model on an infinite lattice, the
average magnetization do not show any phase transition. But, as it
will be shown, the system does exhibit dynamical phase transition.

The homogeneous part of (\ref{2}) can be written as
\begin{equation}\label{16}
\langle\dot s_k\rangle =h_k^l\langle s_l\rangle .
\end{equation}
The eigenvalues and eigenvectors of the operator $h$ satisfy
\begin{align}\label{17}
E\; x_k&= -2x_k+\tanh (2\beta J)(x_{k+1}+x_{k-1}),\quad
k\ne 1,L\nonumber\\
E\; x_1&= -(g_2+g_3)x_1+(g_1-g_2)x_2,\nonumber\\
E\; x_L&= -(h_2+h_3)x_L+(h_1-h_2)x_{L-1},
\end{align}
where the eigenvalue and the eigenvector have been denoted by $E$
and $x$, respectively. The solution to these is
\begin{equation}\label{18}
x_k=a z_1^k+b z_2^k,
\end{equation}
where
\begin{align}\label{20}
-(E+g_2+g_3)(a z_1+b z_2)+(g_1-g_2)(a z_1^2+b z_2^2) &= 0\nonumber\\
-(E+h_2+h_3)(a z_1^L+b z_2^L)+(h_1-h_2)(a z_1^{L-1}+b z_2^{L-1})
&= 0,
\end{align}
and $z_j$'s satisfy
\begin{equation}\label{19}
E=-2+\tanh (2\beta J)(z+z^{-1}).
\end{equation}
So, $z_1z_2=1$. Two cases may occur,

\noindent a) $z_1$ and $z_2$ are phases.

\noindent b) $z_1$ and $z_2$ are both real, not equal to $\pm 1$.
Then the modulus of one of them is less than one, that of the
other is greater than one.

Using (\ref{19}) and  $z_1z_2=1$, one can eliminate $E$, and
arrives at
\begin{align}\label{22}
z^{1-L}[2-g_2-g_3+z(g_1-g_2-\tanh (2\beta J))-z^{-1}\tanh (2\beta
J)] &\nonumber\\ \times[2-h_2-h_3+z(h_1-h_2-\tanh (2\beta
J))-z^{-1}\tanh (2\beta J)]\nonumber&\\ -
z^{L-1}[2-g_2-g_3+z^{-1}(g_1-g_2-\tanh
(2\beta J))-z\tanh (2\beta J)]&\nonumber\\
\times [2-h_2-h_3+z^{-1}(h_1-h_2-\tanh (2\beta J))-z\tanh (2\beta
J)]=0&.
\end{align}
Obviously, $z_j=\pm 1$ satisfies  (\ref{22}). But these solutions
lead to
\begin{equation}\label{28}
x_k=z^k(a +b k).
\end{equation}
And this form for $x_k$, generally does not satisfy the boundary
conditions at $k=1,L$. Equation (\ref{22}) can be written in the
form
\begin{equation}\label{29}
  G(z):=F(z)-F(z^{-1})=0,
\end{equation}
where
\begin{align}\label{30}
F(z):=z^{1-L}\{ 2-g_2-g_3+z[g_1-g_2-\tanh (2\beta
J)]-z^{-1}\tanh(2\beta J)\}\nonumber\\
\times \{2-h_2-h_3+z[h_1-h_2-\tanh (2\beta J)]-z^{-1}\tanh (2\beta
J)\}.
\end{align}
For a phase solution to (\ref{22}), $z=e^{i\vartheta}$, we have
\begin{equation}\label{31}
  E=-2+2\tanh (2\beta J)\cos\vartheta .
\end{equation}
In the thermodynamic limit $(L\to \infty )$, in any neighborhood
of $z=1$ there exist a phase solution to (\ref{22}). The supermum
of the eigenvalues determines the relaxation time toward the
stationary average-density profile. So, if all of the solutions
are phase,
\begin{equation}\label{31b}
  \tau =[-2+2\tanh (2\beta J)]^{-1}
\end{equation}
But, if there exist solutions which are not phases, they should be
real. Consider $z>1$. Then for $L \to \infty$, (\ref{22}) becomes
\begin{align}\label{32}
[2-g_2-g_3+&z^{-1}(g_1-g_2-\tanh (2\beta J))-z\tanh (2\beta J)]\nonumber\\
\times [2-h_2-h_3+&z^{-1}(h_1-h_2-\tanh (2\beta J))-z\tanh (2\beta
J)]=0.
\end{align}
Changing the rates, one may arrive at a situation where the above
equation has a real solution greater than one. The transition
occurs  at the point that this equation has a solution equal to
one. When the system has passed this point, the relaxation time
becomes
\begin{equation}\label{31c}
  \tau =[-2+2(\Lambda +\Lambda^{-1})\tanh (2\beta J)]^{-1},
\end{equation}
where $\Lambda$ is that solution to (\ref{32}), which is greater
than one. (here we have assumed $ J>0$, ferromagnetism. If  $J<0$,
anti-ferromagnetism, $ \Lambda $ is that solution to \Ref{32}
which is less than $-1$.) Putting $z=1$ in (\ref{32}), at least
one of the following equations should hold
\begin{align}\label{34}
  &2[1-\tanh (2\beta J)]-g_2-g_4=0\nonumber \\
  &2[1-\tanh (2\beta J)]-h_2-h_4=0
\end{align}
If the temperature is zero, (\ref{34}) for example gives
$g_2+g_4=0$. Remembering that these parameters are rates, one
arrives at $g_2=g_4=0$. So, at zero temperature, the solution
cannot pass $z=1$. But at any other temperature, $1-\tanh (2\beta
J)$ is positive, and changing the parameters, $g_2+g_4$ can be
made more than or less than $1-\tanh (2\beta J)$.

If one uses the expressions \Ref{gggg} and \Ref{fbar} for $g_i$'s,
then \Ref{34} becomes
\begin{equation}\label{34b}
  -2 \tanh (2\beta J) + \bar f [ \tanh \beta ( J-B_1)+\tanh \beta (
  J+B_1)]=0.
\end{equation}
Putting $\bar f =1$, means that the {\it inertia} of the first
spin against the spin flip is the same as those of the bulk spins.
In this case, however, \Ref{34b} has no solution. That is, there
is no phase transition. In fact, \Ref{34b} has no solution for
$\bar f\leq 1$. For $\bar f >1$, however, it may have a solution.

It is seen that the parameters $g_2$ and $g_4$ (or $h_2$ and
$h_4$) are control parameters of the dynamical phase transition.
The parameters $g_1$ and $g_3$ (or $h_1$ and $h_3$) do not have
any contribution in the dynamical phase transition. The rates
$g_1$ and $g_3$ are the rates of the disappearance of the domain
walls. But we note that the eigenvector corresponding to $z=1$ is
a configuration where all the spins are the same ($s_k\sim
z^k=1$.) It is this configuration which corresponds to the largest
value of $E$, which determines the relaxation time, and in this
configuration, there is no domain wall. The disappearance rate of
this configuration determines the relaxation time towards the
steady state, and $g_1$ and $g_3$ (or $h_1$ and $h_3$) are
irrelevant to this rate. In the particle-vacancy picture, this
means that the rate of change of vacancy to particle is important,
since the configuration corresponding to the maximum value of $E$
is the empty lattice.

This arguments are true for $J>0$, the ferromagnetic case. If
$J<0$, then the relaxation time is determined by the value of $E$
at the smallest possible value of $z$ (which is less than $-1$),
and the transition occurs as $z=-1$ becomes a solution to
\Ref{32}. It is not difficult to see that in this case $g_1+g_3$
(or $h_1+h_3$) determine the phase transition. The reasoning is
the same as above, except that here the configuration determining
the relaxation time is that corresponding to $z=-1$, which means
that the spins are alternating. So, in this configuration there
are no $\uparrow\uparrow$ or $\downarrow\downarrow$ configurations
and $g_2$ and $g_4$ (or $h_2$ and $h_4$) are irrelevant.

\vskip\baselineskip
\noindent {\bf Acknowledgement} The authors would like to thank
Institute for Studies in Theoretical Physics and Mathematics for
partial support.

\end{document}